# Experimental control of mode-competition dynamics in a chaotic multimode semiconductor laser for decision making

Ryugo Iwami,[1,*] Takatomo Mihana,[2] Kazutaka Kanno,[1] Makoto Naruse,[2] and Atsushi Uchida[1,**]

[1]*Department of Information and Computer Sciences, Saitama University, 255 Shimo-okubo, Sakura-ku, Saitama City, Saitama 338-8570, Japan*
[2]*Department of Information Physics and Computing, Graduate School of Information Science and Technology, The University of Tokyo, 7-3-1 Hongo, Bunkyo-ku, Tokyo 113-8656, Japan*
[*]*r.iwami.692@ms.saitama-u.ac.jp*
[**]*auchida@mail.saitama-u.ac.jp*

**Abstract:** Photonic computing has been widely used to accelerate the computational performance in machine learning. Photonic decision-making is a promising approach that uses photonic computing technologies to solve the multi-armed bandit problem based on reinforcement learning. Photonic decision making using chaotic mode competition dynamics has been proposed. However, the experimental conditions for achieving superior decision-making performance have not yet been established. In this study, we experimentally investigate the mode competition dynamics in a chaotic multimode semiconductor laser in the presence of optical feedback and injection. We control chaotic mode-competition dynamics via optical injection, and we found that positive wavelength detuning results in an efficient mode concentration to one of the longitudinal modes with a small optical injection power. We experimentally investigate two-dimensional bifurcation map of the laser dynamics of the total intensity. Complex mixed dynamics are observed in the presence of optical feedback and injection. We experimentally conduct decision making to solve the bandit problem using chaotic mode-competition dynamics. A fast mode concentration property is observed at positive wavelength detuning, which resulted in a fast convergence of the correct decision rate. Our findings could be useful in accelerating the decision-making performance in adaptive optical networks using reinforcement learning.

## 1. Introduction

The demand for large computational resources to support the rapid evolution of artificial-intelligence technologies has increased in recent years. However, the performance of electronic general-purpose computers cannot satisfy the requirements of many calculations owing to the limitations of semiconductor integration technologies, known as the end of Moore's law. To overcome this issue, photonic computing has been widely applied, including photonic artificial neural networks [1], photonic reservoir computing [2–4], coherent Ising machines [5], and nonlinear classification based on silicon photonic circuits [6]. These studies are expected to lead to photonic accelerators that enhance specific computational tasks using photonic technologies [7,8].

Photonic decision making is a promising photonic computing technology that can solve the multi-armed bandit problem (MAB) based on reinforcement learning [9–14]. In MAB, a player maximizes the total reward from multiple slot machines with unknown hit probabilities (an example of multiple choice) [15,16]. These two actions are important for maximizing the total reward. One action, called exploration, searches for the best choice (e.g., a slot machine with the highest hit probability). The other action is called exploitation, which involves finding a choice based on the information obtained from exploration. However, a tradeoff between exploration and exploitation has been reported, in which the total reward is reduced if either exploration or exploitation is biased [16]. The tradeoff between exploration and exploitation is an important issue that must be solved in MAB.

Photonic decision-making for solving MAB with two-slot machines has been reported using chaotic temporal waveforms of a semiconductor laser and a threshold value [9], spontaneous mode dynamics in a semiconductor laser with a ring cavity [11], and coupled semiconductor lasers [12]. To increase the number of slot machines, several schemes have been proposed, such as a comparison of two slot machines in a hierarchical architecture [10], a unidirectionally coupled laser network [13], and the assignment of random bits based on a chaotic semiconductor laser [14]. However, this approach had several limitations. For example, the performance is degraded based on the slot machine arrangement in a hierarchical architecture [10] and decision-making cannot be achieved when the number of slot machines is increased using a coupled laser network [13]. Digital processing is required for random number generation and slot-machine selection [14]. Thus, it is important to propose a method to increase the number of slot machines required for photonic decision making.

Recently, decision making using the dynamics of a multimode semiconductor laser has been proposed by exploring slot machines based on chaotic itinerancy (chaotic mode competition dynamics) [17]. Numerical simulations show that the proposed method requires a smaller number of plays than existing software-based algorithms when the number of slot machines is increased and is useful for solving large-scale MAB. A proof-of-concept experiment has also been reported [17]; however, a detailed experimental investigation is still lacking, and the use of proper dynamics is important to accelerate decision-making performance.

In the dynamics of multimode semiconductor lasers with optical feedback, low-frequency fluctuations (LFF) have been observed numerically and experimentally [18–22]. A bifurcation diagram of the total intensity was experimentally investigated using a multimode semiconductor laser when the feedback power and injection current were changed [23]. In addition, several studies have reported experimental observations of chaotic antiphase dynamics [24] and adaptive mode selection of the maximum modal intensity (called the dominant mode) [25].

Semiconductor lasers have been utilized for photonic reservoir computing [26,27], bandwidth enhancement of chaotic outputs [28], and control of chaotic mode competition dynamics in multimode semiconductor lasers [29]. However, these studies are primarily limited to numerical simulations. It has been reported that semiconductor lasers with optical injection exhibit rich nonlinear dynamics by changing the initial wavelength detuning and the optical injection power [30–33]. It is important to experimentally investigate the conditions for controlling the mode-competition dynamics in a chaotic multimode semiconductor laser because the use of proper dynamics can improve decision-making performance.

In this study, we experimentally investigate the control of mode competition dynamics in a chaotic multimode semiconductor laser in the presence of both optical feedback and injection. We investigate the bifurcation diagram when the optical injection power and initial wavelength detuning are varied. We experimentally evaluate decision-making performance under different conditions of initial wavelength detuning and nonlinear dynamics.

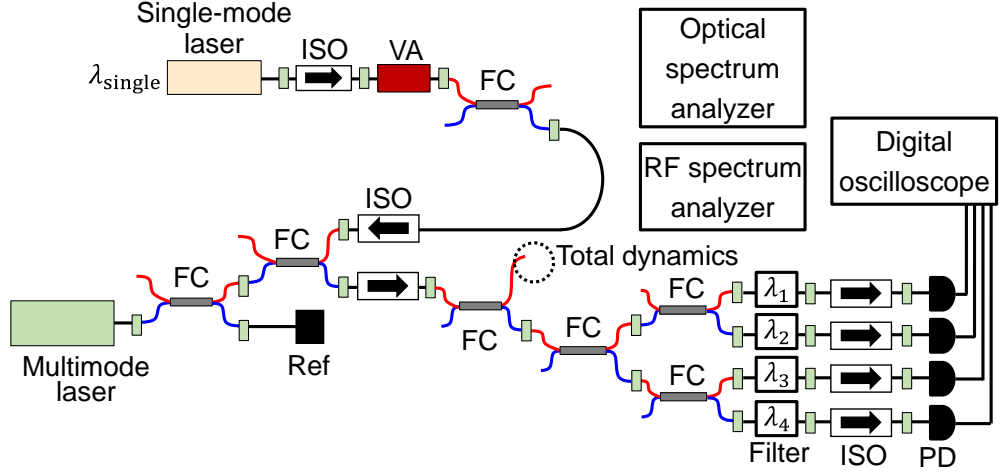

Fig. 1. Experimental setup for multimode semiconductor laser with optical feedback and injection. ISO: isolator, VA: variable attenuator, FC: fiber coupler, Ref: reflector, Filter: wavelength filter, PD: photodetector.

## 2. Dynamics of a chaotic multimode semiconductor laser under optical feedback and injection

### 2.1 Experimental setup

We investigate the dynamics of optical injection into one of the longitudinal modes of a multimode semiconductor laser in the presence of optical feedback. Figure 1 shows the experimental setup for the multimode semiconductor laser with optical feedback and injection. We use a Fabry–Perot multimode semiconductor laser (Anritsu, GF5B5003DLL). The output from the multimode semiconductor laser is reflected by a reflector and reinjected as optical feedback. The dynamics of the multimode semiconductor laser are controlled by optical injection from an external single-mode semiconductor laser (NTT Electronics, KELD1C5GAAA). The output of the single-mode laser is stable, and the wavelength of the single-mode laser is controlled by changing the temperature of the single-mode laser. The injection current of the single-mode laser is set to $I = 57.2$ mA (= 5.3 $I_{th}$). The output of the single-mode laser is injected into the multimode laser. The output of the multimode laser is divided into two outputs to observe the total and modal dynamics. Four-wavelength filters (WL Photonics, IW-WLTF-NE-S-1550-50/0.25-PM-0.9/1.0-FC/APC-USB, FWHM = 0.25 nm) are used to extract the four longitudinal modes. The laser output is converted into an electric signal by a photodetector (Newport, 1544-B, 12 GHz bandwidth), and the temporal waveform is observed using a digital oscilloscope (Tektronix, DPO72304SX, 23 GHz bandwidth, 100 GSample/s). In the experiment, the DC component of the photodetectors is not eliminated to compare the powers of the modal intensities. The RF spectrum of the converted laser output is also observed using an RF spectrum analyzer (Keysight, N9010B, 26.5 GHz bandwidth). Total output in the multimode semiconductor laser (i.e., total intensity) is observed at the fiber indicated by "Total dynamics" in Fig. 1. The optical spectrum of the multimode semiconductor laser is observed using an optical spectrum analyzer (Yokogawa, AQ6370D). Injection current and temperature of the multimode laser are set to $I = 44.73$ mA (= 2.0 $I_{th}$) and $T = 25.00$ °C, respectively. The output power of the multimode laser without optical feedback or injection is 6.9 mW.

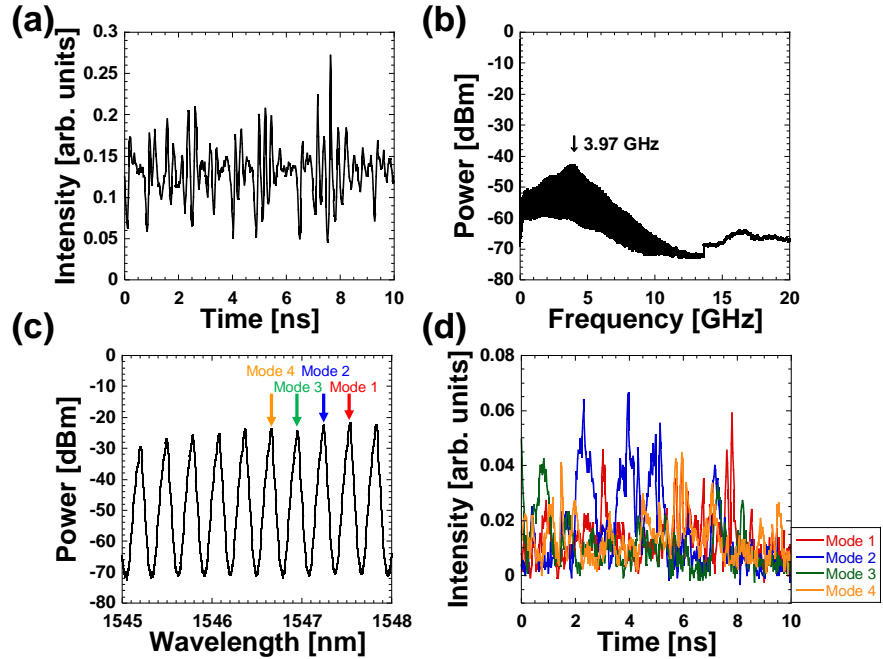

Fig. 2. Dynamics of the multimode semiconductor laser under optical feedback without optical injection. (a) Temporal waveform of total intensity. (b) RF spectrum of total intensity. (c) Optical spectrum of the multimode semiconductor laser. (d) Modal intensities for four modes (mode 1, 2, 3, and 4 are defined in Fig. 2(c)).

*2.2 Modal intensities and optical spectra*

Figure 2 shows the dynamics of a multimode semiconductor laser under optical feedback (without optical injection). Optical feedback power to the multimode laser is estimated as 53 μW, and the total output power of the multimode laser is 6.9 mW. Thus, the power ratio of the feedback light to the total output of the multimode laser is $\kappa_f = 0.0077$. Figure 2(a) shows the temporal waveform of the total intensity, where a chaotic oscillation is observed. Figure 2(b) shows the RF spectrum corresponding to Fig. 2(a). The RF spectrum has smooth and broad peak frequencies, with a maximum peak at a relaxation oscillation frequency of approximately 4 GHz, which is the characteristic frequency of semiconductor lasers [32]. Figure 2(c) shows the optical spectrum of the multimode semiconductor laser with optical feedback. Multiple longitudinal modes are observed in the optical spectra. The longitudinal mode spacing is 0.29 nm (36 GHz in frequency). Four neighboring longitudinal modes with high power are selected and defined as modes 1, 2, 3, and 4, as shown in Fig. 2(c). The wavelengths of mode 1, 2, 3, and 4 are 1547.536, 1547.244, 1546.951, and 1546.661 nm, respectively. The center wavelength of each wavelength filter is set to the wavelength of each longitudinal mode to extract the modal intensities. Figure 2(d) shows the temporal waveforms of the modal intensities for modes 1, 2, 3, and 4 extracted using the four-wavelength filters. Each mode oscillates chaotically, and the temporal waveforms are different. Chaotic mode competition dynamics are observed, that is, one mode oscillates with a large amplitude, and the other modes are suppressed (also known as chaotic antiphase dynamics [24]).

We control the chaotic mode competition dynamics using optical injection. Optical injection from a single-mode semiconductor laser is applied for one of the longitudinal modes (e.g., mode 1) in the multimode laser under the optical feedback of 53 μW ($\kappa_f = 0.0077$). We consider the wavelength of the optical spectral peak of the single-mode semiconductor laser as

$\lambda_{\text{single}}$. The wavelength of the spectral peak of mode 1 in the multimode laser is also defined as $\lambda_{\text{multi},1}$ under optical feedback. The initial wavelength detuning is defined as $\Delta\lambda_{\text{ini}} = \lambda_{\text{multi},1} - \lambda_{\text{single}}$ and we investigate the dynamics for different $\Delta\lambda_{\text{ini}}$. The wavelength detuning is defined to match the sing of the frequency detuning used in the literature [32,33], where the frequency detuning is defined as $\Delta f_{\text{ini}} = f_{\text{drive}} - f_{\text{response}} = f_{\text{single}} - f_{\text{multi},1}$.

The initial wavelength detuning is tuned by varying $\lambda_{\text{single}}$, which is controlled by the temperature of the single-mode laser. For example, the wavelength detuning of $\Delta\lambda_{\text{ini}} = -0.060$ nm is obtained by setting $\lambda_{\text{single}}$ of 1547.596 nm at the temperature of 23.00 °C when $\lambda_{\text{multi},1}$ is fixed at 1547.536 nm. The detuning of $\Delta\lambda_{\text{ini}} = 0.060$ nm is obtained by setting $\lambda_{\text{single}}$ of 1547.476 nm at the temperature of 21.91 °C.

We investigate the optical spectra and modal intensities at different initial wavelength detunings under the optical injection of 350 µW into the multimode laser with optical feedback. The power ratio of the optical injection light of the single-mode laser to the total output of the multimode laser (6.9 mW) is $\kappa_{\text{inj}} = 0.051$. In addition, the power ratio of the optical injection light to the optical feedback light in the multimode laser is $\kappa_{\text{inj}} / \kappa_{\text{f}} = 6.6$.

Figure 3(a) shows the optical spectrum of the multimode laser at $\Delta\lambda_{\text{ini}} = -0.060$ nm. The power of mode 1 increases under optical injection, whereas the powers of the other modes are slightly suppressed. The mode around the wavelength of 1543.426 nm has a higher power than other modes, which corresponds to the original peak wavelength of the multimode laser without optical feedback and injection. Figure 3(b) shows the optical spectrum at $\Delta\lambda_{\text{ini}} = 0.00$ nm. Mode 1 has a high power, and the other modes are suppressed by more than 40 dB, which indicates that the laser power is concentrated in one mode. Figure 3(c) shows the optical spectrum at $\Delta\lambda_{\text{ini}} = +0.060$ nm. Mode 1 has a larger power, and the other modes are suppressed. However, some side modes with high power exist, compared to Fig. 3(b). Thus, the power of the mode is enhanced via optical injection, and the suppression of the other modes is less effective, even at the same optical injection power.

Figure 3(d) shows temporal waveform of the modal intensities for the four modes at $\Delta\lambda_{\text{ini}} = -0.060$ nm. The four modes compete with a small amplitude, and mode 1 has a slightly higher intensity than the other modes. Figure 3(e) shows the temporal waveform of the modal intensities at $\Delta\lambda_{\text{ini}} = 0.00$ nm. Mode 1 is enhanced, and a periodic oscillation with a large amplitude is observed. Figure 3(f) shows the temporal waveform of the modal intensities at $\Delta\lambda_{\text{ini}} = +0.060$ nm. Mode 1 is enhanced, and chaotic oscillations are observed. Therefore, different temporal dynamics are obtained for the three different $\Delta\lambda_{\text{ini}}$, even though the injection power is the same.

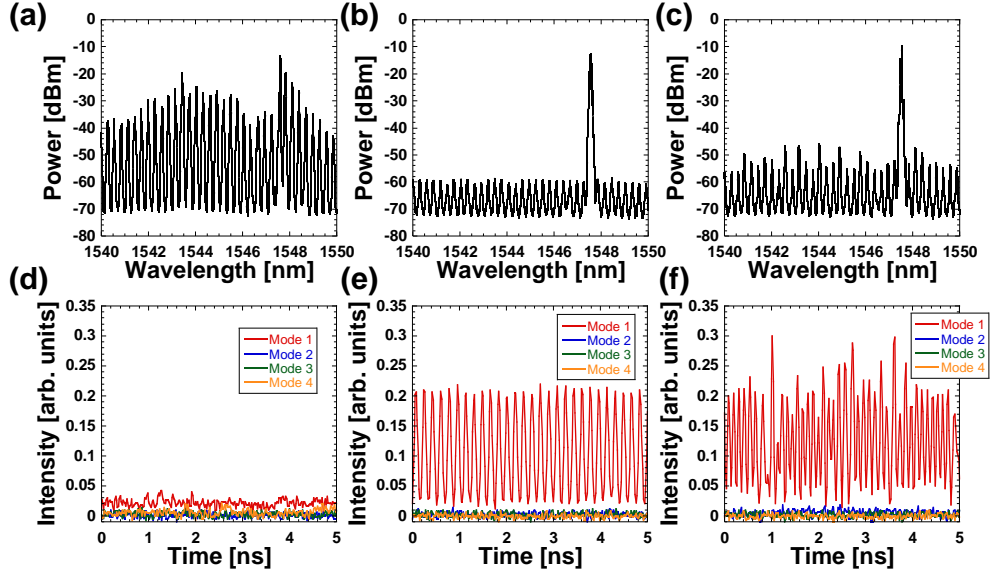

Fig. 3. (a), (b), (c) Optical spectra and (d), (e), (f) temporal waveforms of modal intensities in the multimode semiconductor laser with optical feedback and injection for different initial wavelength detunings. Optical injection power to the multimode laser is estimated to 350 μW ($\kappa_{inj}$ = 0.051). (a), (d) $\Delta\lambda_{ini}$ = −0.060 nm, (b), (e) $\Delta\lambda_{ini}$ = 0.00 nm, (c), (f) $\Delta\lambda_{ini}$ = +0.060 nm.

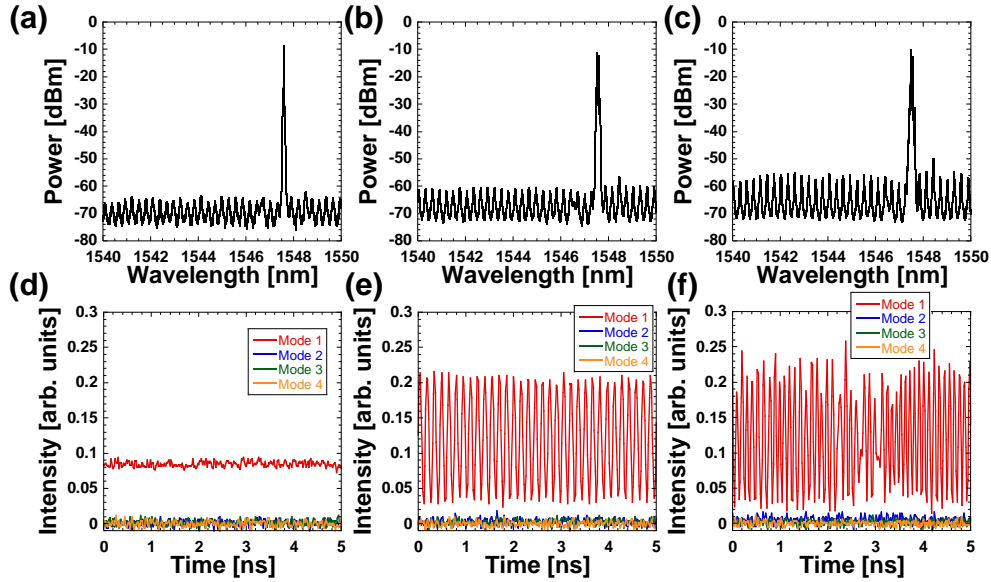

Fig. 4. (a), (b), (c) Optical spectra and (d), (e), (f) temporal waveforms of modal intensities in the multimode semiconductor laser with optical feedback and injection for different initial wavelength detunings. Optical injection power to the multimode laser is estimated to 1030 μW ($\kappa_{inj}$ = 0.15). (a), (d) $\Delta\lambda_{ini}$ = −0.060 nm, (b), (e) $\Delta\lambda_{ini}$ = 0.00 nm, (c), (f) $\Delta\lambda_{ini}$ = +0.060 nm.

Next, we increase the optical injection power to 1030 μW ($\kappa_{inj}$ = 0.15 and $\kappa_{inj} / \kappa_f$ = 19.4) and observe the optical spectra and modal intensities at different initial wavelength detunings.

Figure 4(a) shows the optical spectrum at $\Delta\lambda_{ini}$ = −0.060 nm. The power of mode 1 increases with a large DC component, and the powers of the other modes are suppressed. A sharp peak is observed, unlike the weak optical injection of 350 μW in Fig. 3(a). Figures 4(b) and 4(c) show the optical spectra at $\Delta\lambda_{ini}$ = 0.00 nm and $\Delta\lambda_{ini}$ = +0.060 nm, respectively. Mode 1 has a high power, and the other modes are suppressed, as shown in Figs. 4(b) and 4(c), as well as in Figs. 3(b) and 3(c).

Figure 4(d) shows the temporal waveform of the modal intensities for four modes at $\Delta\lambda_{ini}$ = −0.060 nm. Mode 1 is stabilized and always has the maximum intensity. Figure 4(e) shows the temporal waveform of the modal intensities at $\Delta\lambda_{ini}$ = 0.00 nm. Mode 1 is enhanced and a periodic oscillation with a large amplitude is observed. Figure 4(f) shows the temporal waveform of the modal intensities at $\Delta\lambda_{ini}$ = +0.060 nm. Mode 1 is enhanced and oscillates chaotically. Therefore, different nonlinear dynamics are observed by varying the initial wavelength detuning.

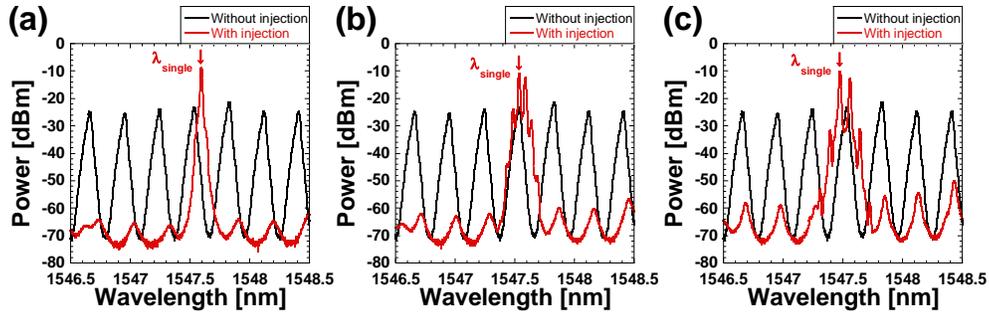

Fig. 5. Optical spectra of the multimode semiconductor laser without and with optical injection (enlarged view of Figs. 4(a), 4(b), and 4(c)). Optical injection power to the multimode laser is estimated to 1030 μW ($\kappa_{inj}$ = 0.15). (a) $\Delta\lambda_{ini}$ = −0.060 nm, (b) $\Delta\lambda_{ini}$ = 0.00 nm, (c) $\Delta\lambda_{ini}$ = +0.060 nm.

Figure 5 shows enlarged view of optical spectra of the multimode laser (corresponding to Figs. 4(a), 4(b), and 4(c)) without and with optical injection for the three different $\Delta\lambda_{ini}$. In Fig. 5(a) of at $\Delta\lambda_{ini}$ = −0.060 nm, one peak of the optical spectrum is obtained with suppression of the other side modes. The peak spectrum of the multimode laser is shifted to a longer wavelength by optical injection (i.e., redshift) and exactly matches that of the single-mode laser, which indicates injection locking (optical frequency matching). The other side modes are also shifted by the optical injection, although the power spectra are very small. The amount of the wavelength shift is 0.06 nm (7.5 GHz in frequency), corresponding to the absolute value of the initial wavelength detuning. This stable locking results in a stable laser output, as shown in Fig. 4(d).

Figure 5(b) shows the case of $\Delta\lambda_{ini}$ = 0.00 nm. Two main peaks are observed in the optical spectrum. The left peak corresponds to the wavelength of the injected single-mode laser, and the right peak corresponds to the wavelength of the multimode laser, which is redshifted by 0.052 nm (6.5 GHz) by optical injection. The two peaks are not matched and injection locking is not achieved at $\Delta\lambda_{ini}$ = 0.00 nm. Other small peaks generated by four-wave mixing are observed at intervals of 0.052 nm. In addition, the other side modes are red-shifted. The existence of these two main peaks induces a periodic temporal waveform, as shown in Fig. 4(e).

Figure 5(c) shows the case of $\Delta\lambda_{ini}$ = +0.060 nm. Two main peaks are obtained with a larger mode spacing than that shown in Fig. 5(b). The left peak corresponds to the wavelength of the injected single-mode laser and the right peak corresponds to the red-shifted wavelength of the multimode laser. The wavelength shift of the multimode laser is 0.024 nm (3.0 GHz) and the mode spacing between the two peaks is 0.084 nm (10.5 GHz). Injection locking is not achieved, and the interaction between the two peaks results in chaotic oscillation of the laser intensity, as shown in Fig. 4(f). Four-wave mixing is observed with several small peaks. The side modes are shifted by 0.024 nm under optical injection. In addition, side-mode suppression at $\Delta\lambda_{ini} =$ +0.060 nm is more effective than the case of $\Delta\lambda_{ini} = -0.060$ nm, as shown in Figs. 3(a) and 3(c).

From these results, we determined that different optical spectra and temporal dynamics can be observed for positive and negative wavelength detuning. For negative-wavelength detuning, injection locking is achieved and a stable temporal output is obtained. For zero- and positive-wavelength detuning, injection locking is not achieved and periodic or chaotic oscillations is observed. This asymmetric behavior results from the linewidth enhancement factor known as the alpha parameter [32,33].

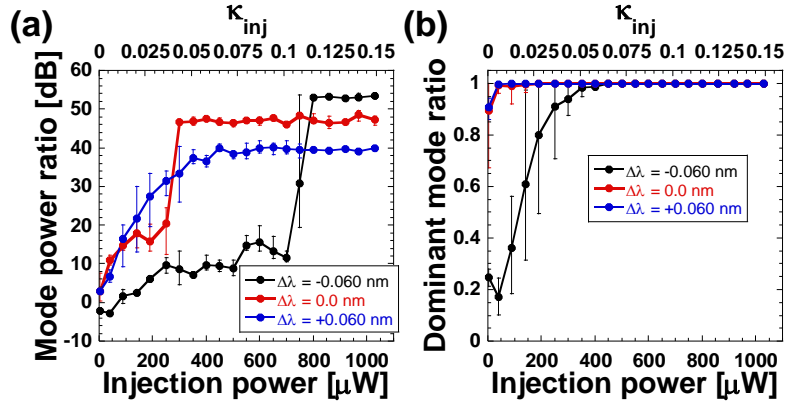

Fig. 6. (a) Mode power ratio (MPR) and (b) dominant mode ratio (DMR) of mode 1 for different initial wavelength detunings when optical injection power is varied. Average value is plotted as a dot, and maximum and minimum values are indicated as error bar. The injected power is adjusted using a variable attenuator, so that the minimum injection power of 3 μW is added even at the lowest injection power (leftmost point in the figure).

*2.3 Mode power ratio and dominant mode ratio*

We measure the interaction among the longitudinal modes using two quantities when the optical injection power for mode 1 is changed at different initial wavelength detunings. The first quantity is the mode power ratio (MPR), which is defined as the difference in the peak power in the optical spectrum as follows:

$$\text{MPR} = P_{\text{main}} - P_{\text{side}} \qquad (1)$$

where $P_{\text{main}}$ is the maximum peak power of mode 1 and $P_{\text{side}}$ is the maximum peak power of the side modes, except for the main mode in the optical spectra. The power of the main mode is defined in the range of 1547.260 to 1547.810 nm to obtain $P_{\text{main}}$; therefore, multiple peaks within mode 1, as shown in Figs. 5(b) and 5(c), are not considered as side modes. For example, in Fig. 5(c), MPR of 39.80 dB is obtained from $P_{\text{main}} = -10.12$ dB at 1547.476 nm and $P_{\text{side}} = -49.92$ dB at 1548.432 nm.

Figure 6(a) shows MPR for the three initial wavelength detunings ($\Delta\lambda_{ini} = -0.060$, 0.00, and +0.060 nm) when the optical injection power for mode 1 is increased. The average MPR is estimated from the optical spectra by varying the injection power. The MPR increases with the injection power. However, the curves are different between the positive and negative detunings (i.e., $\Delta\lambda_{ini} = -0.060$ and +0.060 nm) even though the absolute value of the initial wavelength detuning is the same, which indicates that there are asymmetric characteristics regarding the initial wavelength detuning. The MPR gradually increases with the injection power for positive wavelength detuning. We believe that one of the longitudinal modes can be excited more easily by supplying power from the optical injection light to the longitudinal mode during positive-wavelength detuning. However, in the case of negative-wavelength detuning, the MPR increases rapidly with a large injection power. We consider that mode concentration does not occur under a small injection power. When the injection power reaches the threshold for injection locking, a rapid mode concentration is observed, and a large MPR is obtained. Thus, a larger power is required to achieve a large MPR for negative-wavelength detuning because of injection locking. For a zero-detuning case ($\Delta\lambda_{ini}$ = 0.00 nm), rapid transition of MPR is observed with a small injection power without injection locking.

When we focus on the value of the maximum MPR, it reaches more than 50 dB for the negative wavelength detuning ($\Delta\lambda_{ini}$ = −0.060 nm), which indicates that the modal power is efficiently concentrated on one of the longitudinal modes. This is because all laser powers are concentrated on a sharper spectrum by injection locking, as shown in Fig. 5(a). However, for the positive wavelength detuning ($\Delta\lambda_{ini}$ = +0.060 nm), the maximum MPR is limited to approximately 40 dB. This is because the suppression level of the other modes is lower than that of the negative detuning, as shown in Fig. 4(c), and the laser power is distributed to the coexistence of the two peaks of the injected light and multimode laser. Thus, the asymmetric characteristics appear in the process of mode concentration in the optical spectra, because of the nonzero α parameter of the semiconductor laser [32,33].

The second quantity is the dominant mode ratio (DMR), which represents the ratio of the dominant mode (i.e., the mode with the maximum modal intensity) by comparing the temporal waveforms of the modal intensities [17,25,29]. The DMR for mode $m$ (DMR$_m$) is defined as

$$\text{DMR}_m = \frac{1}{S}\sum_{j=1}^{S} D_m(j) \tag{2}$$

where $S$ denotes the total number of sampled points. $D_m(j)$ is 1 if mode $m$ is the dominant mode at the $j$-th sampled point and 0 otherwise.

Figure 6(b) shows the DMR of mode 1 at three wavelength detunings when the optical injection power of mode 1 is changed. Temporal waveforms of $S = 5 \times 10^5$ points (10 μs in time) for each mode are used to calculate DMR. The temporal waveforms are acquired ten times at each wavelength detuning and injection power, and the average of the DMRs is shown. The DMR increases with the injection power. However, there are asymmetric characteristics in the DMR between the positive and negative detunings. For the positive detuning ($\Delta\lambda_{ini}$ = +0.060 nm), DMR is large at small injection power, and DMR reaches 1 by slightly increasing the injection power. However, for the negative detuning ($\Delta\lambda_{ini}$ = −0.060 nm), DMR increases gradually as the injection power increases. This characteristic agrees with the MPR results shown in Fig. 6(a).

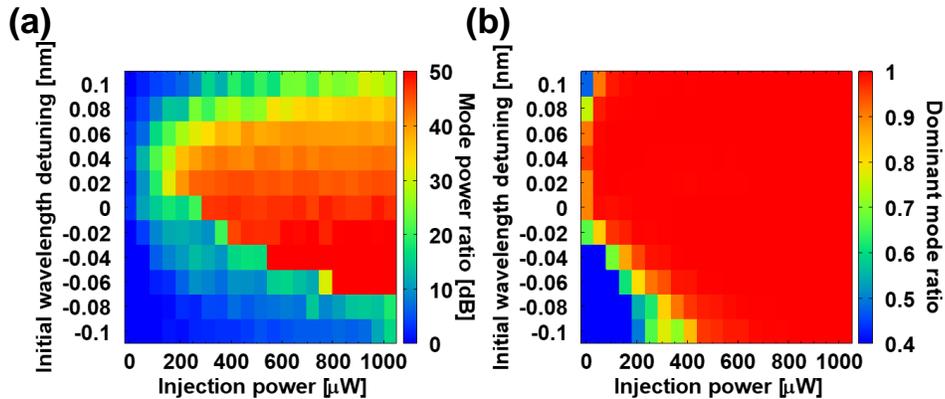

Fig. 7. Two-dimensional maps for (a) MPR and (b) DMR of mode 1 when initial wavelength detuning and optical injection power are changed. The injection power of 1000 μW corresponds to $\kappa_{inj}$ = 0.145.

We investigate the MPR and DMR when the initial wavelength detuning and optical injection power are changed continuously. Figure 7(a) shows the two-dimensional map of MPR. The MPR increases with the injection power for all initial wavelength detunings. However, a higher injection power is required to increase the MPR for negative-wavelength detuning. In addition, large MPR values are achieved for negative-wavelength detuning under strong optical injection. In contrast, for positive wavelength detuning, the MPR rapidly increases with a small injection power. However, the MPR is limited to less than 50 dB.

Figure 7(b) shows a two-dimensional map of the DMR. For negative-wavelength detuning, large optical injection power is required to increase the dominant MPR. In contrast, for positive wavelength detuning, a large DMR is achieved even for a small injection power.

From these results, the asymmetric properties of the MPR and DMR are evident in terms of optical wavelength detuning. More injection power is required for negative wavelength detuning because of the injection locking condition, whereas a small injection power results in an effective mode concentration for positive wavelength detuning without using injection locking. However, larger values of the MPR and DMR are obtained for negative detuning than for positive detuning. The use of injection locking at negative detunings is not always required to improve the performance of some applications, such as decision-making, as described in Section 3.

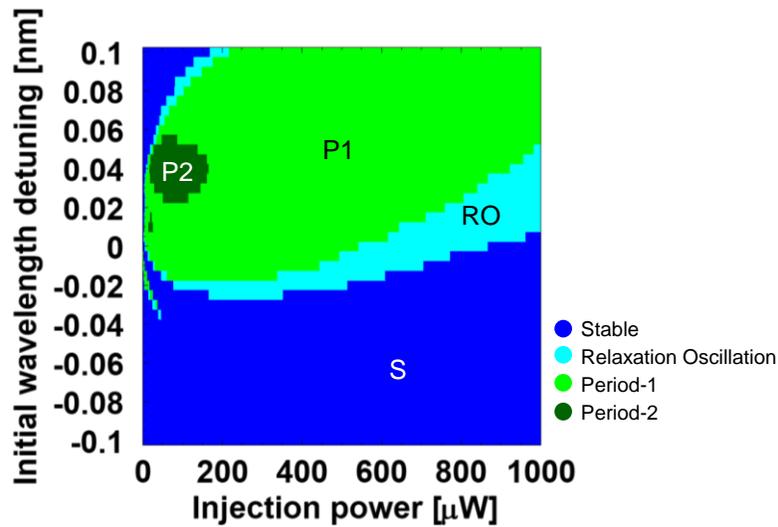

Fig. 8. Two-dimensional bifurcation diagram of the dynamics in the total intensity when the initial wavelength detuning (vertical axis) and optical injection power (horizontal axis) are continuously changed without optical feedback. The injection power of 1000 μW corresponds to $\kappa_{inj} = 0.145$.

### 2.4 Two-dimensional bifurcation diagram

We experimentally investigate the two-dimensional bifurcation diagram of the total intensity dynamics when the initial wavelength detuning and optical injection power change continuously. Temporal waveforms and RF spectra of the total intensity are observed when the injection power increases under a fixed initial wavelength detuning, and the bifurcation points of the laser dynamics are recorded. A two-dimensional bifurcation map is created by repeating this procedure and changing the initial wavelength detuning.

Figure 8 shows a two-dimensional bifurcation diagram of a multimode semiconductor laser without optical feedback under optical injection. The horizontal axis represents the injection power, and the vertical axis represents the initial wavelength detuning. The injection power is changed from 0 to 1000 μW (0.0 to 0.145 in $\kappa_{inj}$). Note that $\lambda_{multi,1}$ is defined as the wavelength of mode 1 in the absence of optical feedback. The total intensity of the multimode laser shows a stable output when the injection power is increased for negative-wavelength detuning. However, for positive wavelength detuning, period-1 oscillations are observed in a wide range of positive wavelength detuning and period-2 oscillations are observed inside the region of period-1 oscillations. Relaxation oscillations are observed between the regions of period-1 oscillations and stable outputs. Period doubling and chaotic oscillations are observed in a single-mode semiconductor laser with optical injection when the injection power and initial wavelength detuning vary [30,31,33]. We observe a small region of period doubling and chaotic oscillations inside the region of period-2 oscillations.

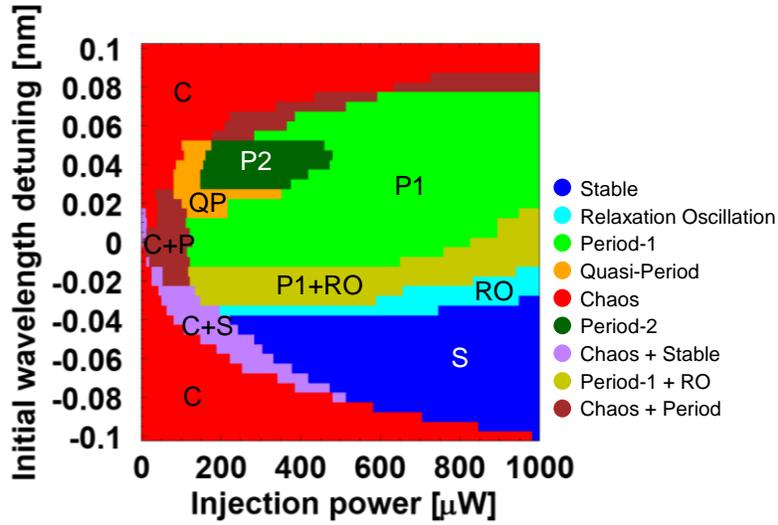

Fig. 9. Two-dimensional bifurcation diagram of the dynamics in the total intensity when the initial wavelength detuning (vertical axis) and optical injection power (horizontal axis) are continuously changed under optical feedback of 14 μW ($\kappa_f$ = 0.0020). The injection power of 1000 μW corresponds to $\kappa_{inj}$ = 0.145.

We also investigate two-dimensional bifurcation diagram of the total intensity dynamics under weak optical feedback power. Figure 9 shows the two-dimensional bifurcation diagram when the optical feedback power is set to 14 μW ($\kappa_f$ = 0.0020). Additionally, the injection power is changed from 0 to 1000 μW (0.0 to 0.145 in $\kappa_{inj}$). Note that $\lambda_{multi,1}$ is defined as the wavelength of mode 1 under optical feedback and without optical injection. Stable outputs (blue) are observed for a wide range of negative wavelength detunings. The stabilization of the chaotic oscillations of the multimode semiconductor laser results from injection locking, in which the wavelength of the multimode laser matches that of the injected light of the stable single-mode semiconductor laser. The injection-locking region is observed for negative wavelength detuning, similar to the case of single-mode semiconductor lasers [32]. Period-1 oscillations (light green) are observed over a wide range of positive wavelength detunings. Period-2 and quasi-periodic oscillations are also observed. Chaotic oscillations (red) are observed for large absolute values of wavelength detuning under weak optical feedback. Quasi-periodic oscillations are determined from multiple discrete peaks with irregular frequency intervals in the RF spectra, and chaotic oscillations are determined from the smooth and broad frequency spectra.

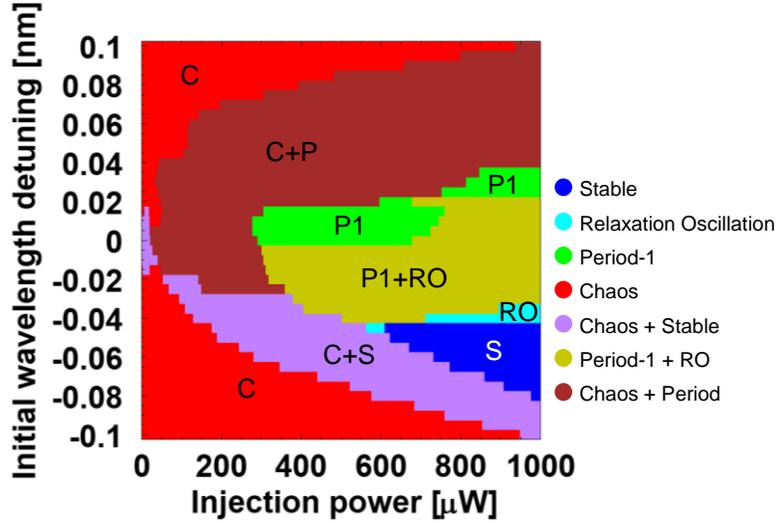

Fig. 10. Two-dimensional bifurcation diagram of the dynamics in the total intensity when the initial wavelength detuning (vertical axis) and optical injection power (horizontal axis) are continuously changed under optical feedback power of 53 μW ($\kappa_f = 0.0077$). The injection power of 1000 μW corresponds to $\kappa_{inj} = 0.145$.

Figure 10 shows the two-dimensional bifurcation diagram under strong optical feedback power of 53 μW ($\kappa_f = 0.0077$) and with optical injection. Additionally, the injection power is changed from 0 to 1000 μW (0.0 to 0.145 in $\kappa_{inj}$). Bifurcation from chaos (red) to other dynamics occur when the optical injection power is increased. The dynamics are stabilized (blue region) in a narrower range of negative wavelength detunings and a large injection power compared to Fig. 9. In contrast, periodic-1 oscillations (light-green region) are observed near zero and with positive wavelength detuning. This is because of the coexistence of the peaks of the injected light and the mode of the multimode laser without injection locking, as observed in the optical spectrum in Fig. 5(b), and the beat frequency of the two peak spectra appearing as periodic oscillations. In addition, mixed-dynamics regions are widely distributed between the chaotic, stable, and period-1 regions. However, the bifurcation diagram is similar to that shown in Fig. 9, although the injection powers at the bifurcation points differ. It is worth noting that mixed dynamics are more likely to be observed under strong optical feedback power.

We speculate that the complex mixed dynamics (e.g., chaotic and periodic oscillations) in Figs. 9 and 10 result from phase fluctuations between the optical injection and feedback light. The coexistence of optical injection and feedback light results in interference within the laser cavity, and phase fluctuations would change the dynamics over time. In fact, we have found that the temporal dynamics are strongly influenced by changes in the phase fluctuations and small-wavelength detuning corresponding to the inverse of the round-trip time in the external cavity of the optical feedback [29]. We consider that the fluctuations of the optical linewidth in single-mode semiconductor lasers and phase fluctuations of the optical feedback in multimode semiconductor lasers may have results in the mixed dynamics observed in the experiment shown in Fig. 10.

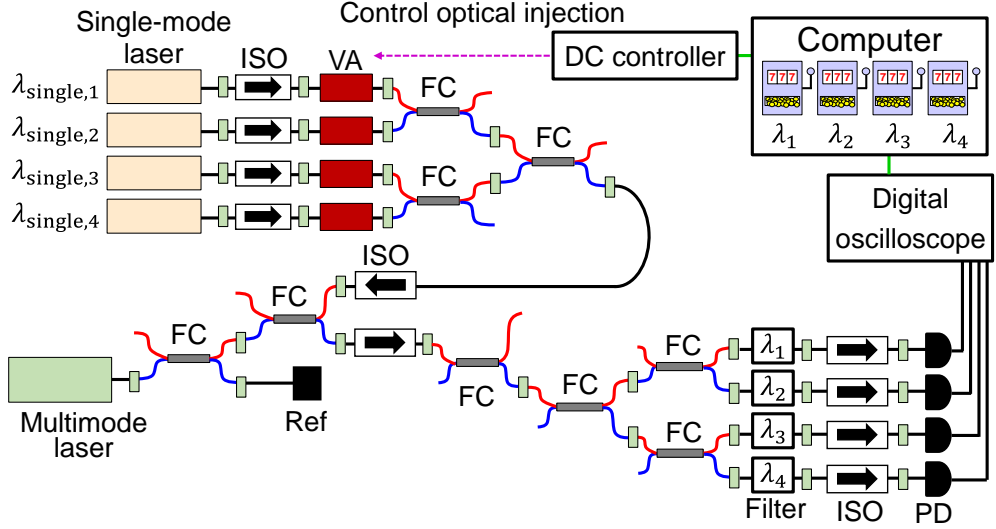

Fig. 11. Experimental setup for decision making using the multimode semiconductor laser with optical feedback and injection. ISO: isolator, VA: variable attenuator, FC: fiber coupler, Ref: reflector, Filter: wavelength filter, PD: photodetector. Four modes in the multimode semiconductor laser are assigned to four slot machines. Result of slot machine selection is emulated in a computer. Optical injection power from four single mode lasers is controlled based on the result of slot machine selection.

## 3. Experiment for decision making

### 3.1 Experimental setup for decision making

In this section, we experimentally investigate the decision-making for solving the MAB using a multimode semiconductor laser with optical feedback and injection. Figure 11 shows the experimental setup for decision making using a multimode semiconductor laser. Each longitudinal mode is assigned to a slot machine, and the slot machine corresponding to the dominant mode is selected for decision-making [17]. Four single-mode lasers are used to control the modal intensity dynamics of the four modes. The dominant mode changes spontaneously with chaotic mode competition dynamics, which are used for the exploration of the slot machine with the maximum hit probability. The dynamics are controlled via optical injection, which is used for exploitation in MAB. Four slot machines are emulated on a computer, and online decision-making is performed experimentally. The temporal waveform of each mode is obtained for 1000 points using an oscilloscope and transferred to a computer. On a computer, modal intensities at the sampled points in the temporal waveforms are compared, and the mode with the maximum intensity is determined to be the dominant mode. Subsequently, a slot machine corresponding to the dominant mode is selected. The result of the slot-machine selection (hit or miss) is obtained according to the hit probability of the selected slot machine in the computer, and the optical injection power from the single-mode lasers is controlled according to the result (see the Appendix for the algorithm). The injection power from the single-mode semiconductor laser can be adjusted by controlling the voltage of the variable electric attenuators. A computer, oscilloscope, and voltage controller are used to control the injection power and are connected by LAN cables for online control. The single-mode semiconductor laser $m$ is used to excite mode $m$ in the multimode semiconductor laser, and its peak wavelength is defined as $\lambda_{single,m}$. Mode $m$ is excited by adjusting $\lambda_{single,m}$ close to the peak wavelength of mode $m$ ($\lambda_{multi,m}$) in the presence of optical feedback. The initial wavelength detuning for mode $m$ is defined as $\Delta\lambda_{ini,m} = \lambda_{multi,m} - \lambda_{single,m}$ and the initial wavelength detunings for four mode are set to the same value for decision making. We conduct decision making under the optical feedback power of 53 μW ($\kappa_f = 0.0077$) in the multimode semiconductor laser.

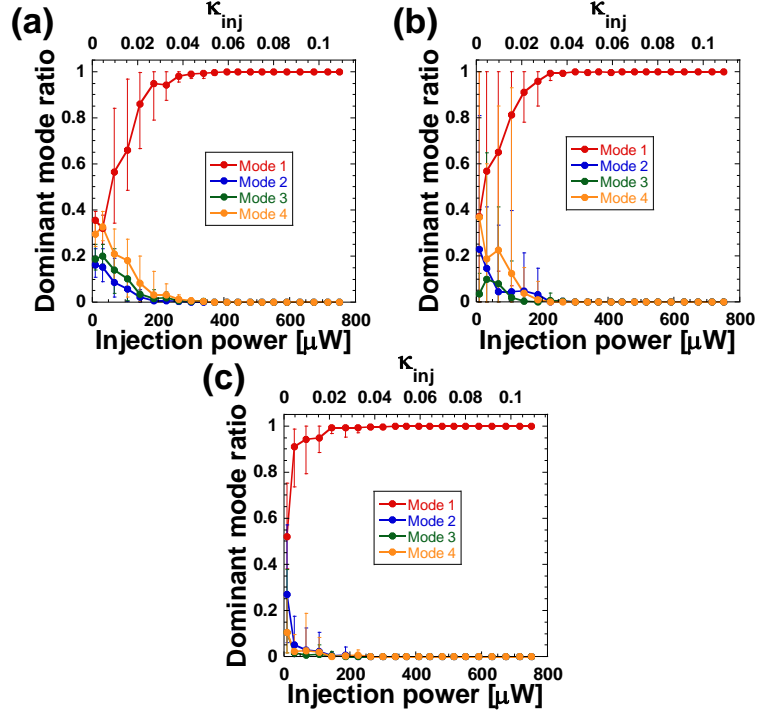

Fig. 12. DMRs for different wavelength detunings when optical injection power for mode 1 is increased in the presence of optical injection for four modes. Optical injection power for mode 2,3,4 is fixed at 9 μW ($\kappa_{inj,2,3,4} = 0.001$). (a) $\Delta\lambda_{ini,m} = -0.060$ nm, (b) $\Delta\lambda_{ini,m} = 0.00$ nm, (c) $\Delta\lambda_{ini,m} = +0.060$ nm.

*3.2 Dominant mode ratios and decision-making performance*

In this section, we focus on three different wavelength detunings of $\Delta\lambda_{ini,m} = -0.060$, 0.00, and +0.060 nm, which show stable output, periodic oscillation, and chaotic oscillation when optical injection power is increased (see Figs. 3 and 4). We conduct decision-making at three-wavelength detunings when we set the same initial wavelength detunings for all modes. We investigate the DMR when the injection power for mode 1 increases in the presence of optical injection. Figure 12 shows the DMRs for the four modes when the injection power for mode 1 is changed. The injection power for mode 2, 3, and 4 are fixed at 9 μW ($\kappa_{inj,2,3,4} = 0.001$). DMR is obtained from the temporal waveforms of $5 \times 10^5$ points (10 μs in time) for each mode. The average of ten DMRs is shown, and their maximum and minimum values are indicated by error bars.

Figures 12(a), 12(b), and 12(c) show DMRs for the wavelength detunings of $\Delta\lambda_{ini,m} = -0.060$, 0.00, and +0.060 nm, respectively. In all cases, the DMR of mode 1 increases as the injection power increases, and the dominant mode could be controlled. Comparing Figs. 12(a) 12(b), the average DMRs appear to be similar. However, the large error bars at low injection powers in Fig. 12(b) show that the DMRs fluctuate between 0 and 1, and that the DMR changes significantly over time. In fact, one of the modes has the maximum intensity for more than 10 μs with periodic oscillations, and the dominant mode changes slowly. This indicates that mode competition occurs at a time scale slower than the GHz chaotic oscillations. In the case of $\Delta\lambda_{ini,m} = +0.060$ nm in Fig. 12(c), DMR increases rapidly with less injection power than the cases in

Figs. 12(a) and 12(b). In addition, the DMR for mode 1 is larger than those of the other modes under a small injection power, as shown in Fig. 12(c). This rapid convergence of the DMR may result from the coexistence of the injected and original modes without injection locking at positive wavelength detunings.

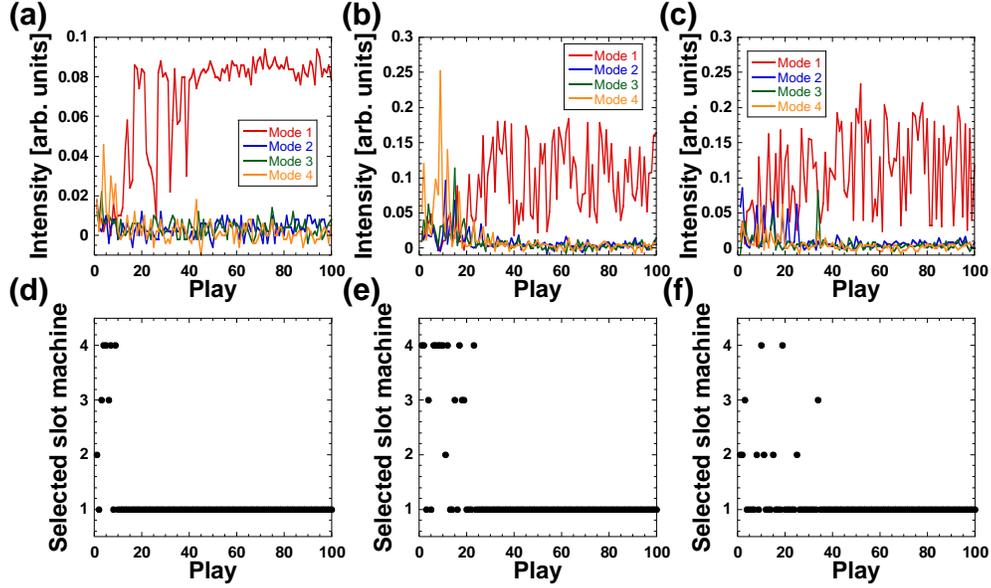

Fig. 13. Results of decision making when the hit probability of slot machine 1 is set to 0.7 and hit probabilities of slot machines 2, 3, and 4 are set to 0.3. (a), (b), (c) modal intensities (d), ©, (f) slot machine selection as a function of the number of plays. (a), (d) $\Delta\lambda_{\text{ini},m} = -0.060$ nm, (b), © $\Delta\lambda_{\text{ini},m} = 0.00$ nm, (c), (f) $\Delta\lambda_{\text{ini},m} = +0.060$ nm.

Next, we perform decision making under different wavelength detunings when the hit probability of slot machine 1 is set to 0.7, and the hit probabilities of slot machines 2, 3, and 4 are set to 0.3. Figure 13 shows the modal intensities for the four modes (Figs. 13(a), 13(b), and 13(c)) and the corresponding slot machine selection (Figure 13(d), 13©, and 13(f)) as a function of the number of plays. Figures 13(a) and 13(d) show modal intensities and slot machine selection for each play at $\Delta\lambda_{\text{ini},m} = -0.060$ nm. The dominant mode in the four modes is changed up to approximately the 10$^{\text{th}}$ play, as shown in Fig. 13(a), and the slot machines are randomly selected, as shown in Fig. 13(d). In contrast, the modal intensity of mode 1 has the largest intensity and is stabilized at a value of 0.08 in Fig. 13(a). As the number of plays increases, slot machine 1 is continuously selected in Fig. 13(d) after the 10$^{\text{th}}$ play. Figures 13(b) and 13© show the modal intensities and slot machine selection for each play at $\Delta\lambda_{\text{ini},m} = 0.00$ nm, and Figures 13(c) and 13(f) show the case at $\Delta\lambda_{\text{ini},m} = +0.060$ nm. The four modes are competed when the number of plays is small and the slot machines are selected randomly. Mode 1 is enhanced as the number of plays increases; however, mode 1 oscillates irregularly, unlike that in Fig. 13(a). Mode 1 has the largest modal intensity compared with the other modes, which allows slot machine 1 to be continuously selected.

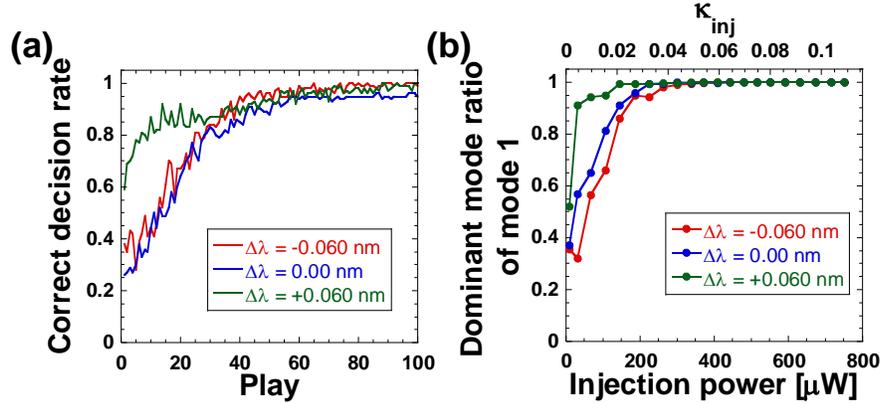

Fig. 14. (a) Correct decision rates for different wavelength detunings when the hit probability of slot machine 1 is set to 0.7 and hit probabilities of slot machine 2, 3, and 4 are set to 0.3. (b) DMRs of mode 1 when optical injection power for mode 1 is increased in the presence of optical injection for four modes. Average values of DMRs of mode 1 in Fig. 12 are plotted.

To evaluate the decision-making performance, we perform multiple cycles of decision-making and evaluate the correct decision rate. Correct decision rate is defined as the rate at which the slot machine with the highest hit probability is selected for each play when multiple decision-making cycles are performed. The correct decision rate for $t$-th play is described as follows [9,10].

$$CDR(t) = \frac{1}{N}\sum_{i=1}^{N} C(i,t) \qquad (3)$$

where $N$ is the total number of decision-making cycles ($N = 100$), $C(i, t)$ is a function that takes the value of 1 if the slot machine with the highest hit probability is selected at $t$-th play in $i$-th cycle, and 0 otherwise.

Figure 14(a) shows the correct decision rates at the three different wavelength detunings of $\Delta\lambda_{\text{ini},m} = -0.060$, $0.00$, and $+0.060$ nm. For all cases, the correct decision rate increases as the number of plays increases, and the slot machine with the highest hit probability can be selected correctly as a result of the exploration. The curves of the correct decision rates for $\Delta\lambda_{\text{ini},m} = -0.060$ nm and $\Delta\lambda_{\text{ini},m} = 0.00$ nm look similar. However, the curve of the correct decision rate for $\Delta\lambda_{\text{ini},m} = +0.060$ nm saturates faster than those for the other wavelength detunings.

We also compare the DMRs when the optical injection power for mode 1 is increased during the decision-making process. Figure 14(b) shows the DMRs of mode 1 for different wavelength detunings (the average DMRs of mode 1 are replotted in Fig. 12(a) and (c)). DMR for $\Delta\lambda_{\text{ini},m} = +0.060$ nm increases with less injection power, and DMRs for $\Delta\lambda_{\text{ini},m} = -0.060$ nm and $\Delta\lambda_{\text{ini},m} = 0.00$ nm are close to each other. Comparing Figs. 14(a) and 14(b), the fast convergence of the DMR corresponds to the fast convergence of the correct decision rate. Correct decision rate increases rapidly at $\Delta\lambda_{\text{ini},m} = +0.060$ nm in Fig. 14(a), because DMR saturates rapidly in Fig. 14(b), which accelerates the selection of the best slot machine. Therefore, the fast mode concentration property results in fast decision-making.

## 4. Discussions

In single-mode semiconductor lasers, injection locking can be achieved by negative-wavelength detuning and is used to match the optical wavelength [32]. We found that fast-mode concentration can be achieved without injection locking in positive-wavelength detunings. Therefore, positive wavelength detuning plays an important role in achieving effective mode concentration in a multimode semiconductor laser. We also observed the complex mixed dynamics in the presence of optical injection and feedback light. We speculate that these mixed dynamics result from phase fluctuations between the optical injection and feedback light.

In this decision-making experiment, each play of the slot machine requires 0.3 s because the slot machine selection in the oscilloscope and control of optical injection are performed electrically in our experiment. Therefore, DMR has a significant impact on decision making and no significant difference in correct decision rates appears for $\Delta\lambda_{\text{ini},m} = -0.060$ nm and $\Delta\lambda_{\text{ini},m} = 0.00$ nm. However, if decision-making is conducted at higher speeds, fast dynamics are expected to have an impact on decision-making performance. The relationship between high-speed laser dynamics and decision-making performance is important for improving the performance of ultrafast photonic computing, and will be addressed in our future work.

## 5. Conclusions

We experimentally investigated the control of the mode-competition dynamics in a chaotic multimode semiconductor laser. The dynamics of optical injection for one of the longitudinal modes in the multimode laser were investigated, and it was found that the laser power was focused on one longitudinal mode with a smaller optical injection power when the initial wavelength detuning was positive. We experimentally investigated the two-dimensional bifurcation diagrams of the total intensity dynamics when the initial wavelength detuning and injection power were continuously changed. We observed complex dynamics in the bifurcation diagram, including mixed dynamics. The region of mixed dynamics was enhanced, and the regions of periodic-1 oscillation and stable output were degenerated by increasing the optical feedback power.

We experimentally conducted decision making for four slot machines using four longitudinal modes. The slot machine corresponding to the dominant mode was selected, and the chaotic mode competition dynamics were controlled according to the slot machine selection results. Decision making was conducted at different wavelength detunings (negative, zero, and positive), which exhibited stable output, periodic oscillation, and chaotic oscillation as the injection power increased. Correct decision-making could be achieved, and positive wavelength detuning resulted in faster convergence of the correct decision rate because the power was concentrated in one of the longitudinal modes with a smaller injection power. The fast-mode concentration property at positive wavelength detuning could be useful for accelerating effective decision-making.

## 6. Appendix

We explain the algorithm of the tug-of-war method for solving the MAB, which is inspired by the behavior of amoebas [34,35]. To control the optical injection power, we change $m$-th evaluation value $X_m(t)$ at $t$-th play in the tug-of-war method, which corresponds to the evaluation of slot machine $m$ at $t$-th play, as follows [17,34,35]. The initial value is set as $X_m(0) = 0$. If slot machine $s$ is selected and a hit is obtained at $t$-th play, $X_m(t)$ of each slot machine $m$ is changed as follows:

$$X_m(t) = \begin{cases} X_m(t-1) + \Delta(t) & (m = s) \\ X_m(t-1) - \dfrac{\Delta(t)}{M-1} & (m \neq s) \end{cases} \quad (4)$$

where $\Delta(t)$ is the amount of change for a hit. $M$ is the number of slot machines; $M = 4$ is used in this study. If slot machine $s$ is selected and a miss is obtained at $t$-th play, $X_m(t)$ of each slot machine $m$ is changed as follows:

$$X_m(t) = \begin{cases} X_m(t-1) - \Omega(t) & (m = s) \\ X_m(t-1) + \dfrac{\Omega(t)}{M-1} & (m \neq s) \end{cases} \quad (5)$$

where $\Omega(t)$ is the amount of change for a miss. $\Delta(t)$ and $\Omega(t)$ are obtained using the estimated hit probabilities as follows [13,17].

$$\Delta(t) = 2 - [\hat{P}_{\text{top1}}(t) + \hat{P}_{\text{top2}}(t)] \quad (6)$$
$$\Omega(t) = \hat{P}_{\text{top1}}(t) + \hat{P}_{\text{top2}}(t) \quad (7)$$

where $\hat{P}_{\text{top1}}(t)$ and $\hat{P}_{\text{top2}}(t)$ are the largest and second-largest estimates of the hit probability for slot machine $m$ of $\hat{P}_m(t)$ up to $t$th play, respectively. The hit probability $\hat{P}_m(t)$ for slot machine $m$ up to $t$-th play is estimated based on the results of the slot machine plays as follows [17]:

$$\hat{P}_m(t) = \begin{cases} \dfrac{R_m(t)}{S_m(t) + 1} & (S_m(t) \neq 0) \\ P_{\text{unknown}}(t) & (S_m(t) = 0) \end{cases} \quad (8)$$

where $R_m(t)$ is the number of hits for slot machine $m$ up to $t$-th play, and $S_m(t)$ is the number of plays for slot machine $m$ up to $t$-th play. $P_{\text{unknown}}(t)$ is used to estimate the hit probability of a slot machine that has never been selected ($S_m(t) = 0$). We set $P_{\text{unknown}}(t)$ to the maximum value of $\hat{P}_m(t)$ at $t$-th play, i.e., $\hat{P}_{\text{top1}}(t)$ is used to facilitate exploration [17]. Based on $X_m(t)$ calculated after $t$-th play, the injection power ratio of injected light for mode $m$ ($\kappa_{\text{inj},m}$) is changed for the next play. The injection power ratio $\kappa_{\text{inj},m}$ is controlled as follows.

$$\kappa_{inj,m} = \begin{cases} \kappa_{\text{inj,max}} & (k \text{ int}(X_m(t)) \geq \kappa_{\text{inj,max}}) \\ \kappa_{\text{inj,min}} & (k \text{ int}(X_m(t)) \leq \kappa_{\text{inj,min}}) \\ k \text{ int}(X_m(t)) & (\text{othewise}) \end{cases} \quad (9)$$

where int($a$) is a function that converts the value $a$ to an integer (if $a$ is a positive number, $a$ is rounded down after the decimal point). $\kappa_{\text{inj,max}}$ and $\kappa_{\text{inj,min}}$ are the maximum and minimum injection power ratio of injected light, respectively. $k$ is the scaling coefficient for converting $X_m(t)$ into the injection power ratio. The injection power ratio of the injected light is then determined based on the value of $k$ int($X_m(t)$) and controlled such that it did not exceed the maximum and minimum values. We set variables to $\kappa_{\text{inj,max}} = 0.11$, $\kappa_{\text{inj,min}} = 0.001$, and $k = 0.0054$. The minimum value of the injected light has been set to zero in our numerical simulation [17]; however, it is set to 0.011 in the experiment. We also adjust the maximum injection power for each mode in the multimode semiconductor laser from single-mode lasers to 750 μW.

**Funding.** Grants-in-Aid for Scientific Research from the Japan Society for the Promotion of Science (JSPS KAKENHI, Grant No. JP19H00868, JP20K15185, and JP22H05195).

**Disclosures.** The authors declare no conflicts of interests.

**Data availability.** Data underlying the results presented in this paper are not publicly available at this time but may be obtained from the authors upon reasonable request.

### References

1. Y. Shen, N. C. Harris, S. Skirlo, M. Prabhu, T. Baehr-Jones, M. Hochberg, X. Sun, S. Zhao, H. Larochelle, D. Englund, and M. Soljačić, "Deep learning with coherent nanophotonic circuits," Nat. Photonics **11**(7), 441–446 (2017).
2. L. Larger, M. C. Soriano, D. Brunner, L. Appeltant, J. M. Gutíerrez, L. Pesquera, C. R. Mirasso, and I. Fischer, "Photonic information processing beyond Turing: An optoelectronic implementation of reservoir computing," Opt. Express **20**(3), 3241–3249 (2012).
3. D. Brunner, M. C. Soriano, C. R. Mirasso, and I. Fischer, "Parallel photonic information processing at gigabyte per second data rates using transient states," Nat. Commun. **4**, 1364 (2013).
4. K. Takano, C. Sugano, M. Inubushi, K. Yoshimura, S. Sunada, K. Kanno, and A. Uchida, "Compact reservoir computing with a photonic integrated circuit," Opt. Express **26**(22), 29424–29439 (2018).
5. T. Inagaki, Y. Haribara, K. Igarashi, T. Sonobe, S. Tamate, T. Honjo, A. Marandi, P. L. McMahon, T. Umeki, K. Enbutsu, O. Tadanaga, H. Takenouchi, K. Aihara, K. I. Kawarabayashi, K. Inoue, S. Utsunomiya, and H. Takesue, "A coherent Ising machine for 2000-node optimization problems," Science **354**(6312), 603–606 (2016).

6. G. Cong, N. Yamamoto, T. Inoue, Y. Maegami, M. Ohno, S. Kita, S. Namiki, and K. Yamada, "On-chip bacterial foraging training in silicon photonic circuits for projection-enabled nonlinear classification," Nat. Commun. **13**, 3261 (2022).
7. K. Kitayama, M. Notomi, M. Naruse, K. Inoue, S. Kawakami, and A. Uchida, "Novel frontier of photonics for data processing—Photonic accelerator," APL Photonics, **4**(9), 090901 (2019).
8. B. J. Shastri, A. N. Tait, T. F. de Lima, W. H. P. Pernice, H. Bhaskaran, C. D. Wright, and P. R. Prucnal, "Photonics for artificial intelligence and neuromorphic computing," Nat. Photonics **15**(2) 102–114 (2021).
9. M. Naruse, Y. Terashima, A. Uchida, and S. J. Kim, "Ultrafast photonic reinforcement learning based on laser chaos," Sci. Rep. **7**, 8772 (2017).
10. M. Naruse, T. Mihana, H. Hori, H. Saigo, K. Okamura, M. Hasegawa, and A. Uchida, "Scalable photonic reinforcement learning by time-division multiplexing of laser chaos," Sci. Rep. **8**, 10890, (2018).
11. R. Homma, S. Kochi, T. Niiyama, T. Mihana, Y. Mitsui, K. Kanno, A. Uchida, M. Naruse, and S. Sunada, "On-chip photonic decision maker using spontaneous mode switching in a ring laser," Sci. Rep. **9**, 9429 (2019).
12. T. Mihana, Y. Mitsui, M. Takabayashi, K. Kanno, S. Sunada, M. Naruse, and A. Uchida, "Decision making for the multi-armed bandit problem using lag synchronization of chaos in mutually coupled semiconductor lasers," Opt. Express **27**(19), 26989–27008 (2019).
13. T. Mihana, K. Fujii, K. Kanno, M. Naruse, and A. Uchida, "Laser network decision making by lag synchronization of chaos in a ring configuration," Opt. Express **28**(26), 40112–40130 (2020).
14. J. Peng, N. Jiang, A. Zhao, S. Liu, Y. Zhang, K. Qiu, and Q. Zhang, "Photonic decision-making for arbitrary-number-armed bandit problem utilizing parallel chaos generation," Opt. Express **29**(16), 25290-25301 (2021).
15. H. Robbins, "Some aspects of the sequential design of experiments," Bull. Am. Math. Soc. **58**(5), 527–535 (1952).
16. R. S. Sutton and A. G. Barto, *Reinforcement Learning: An Introduction*, ed. 2 (The MIT Press, 2018).
17. R. Iwami, T. Mihana, K. Kanno, S. Sunada, M. Naruse, and A. Uchida, "Controlling chaotic itinerancy in laser dynamics for reinforcement learning," Sci. Adv. **8**(49), eabn8325 (2022).
18. G. Vaschenko, M. Giudici, J. J. Rocca, C. S. Menoni, J. R. Tredicce, and S. Balle, "Temporal dynamics of semiconductor lasers with optical feedback," Phys. Rev. Lett. **81**(25), 5536–5539 (1998).
19. G. Huyet, S. Balle, M. Giudici, C. Green, G. Giacomelli, and J. R. Tredicce, "Low frequency fluctuations and multimode operation of a semiconductor laser with optical feedback," Opt. Commun. **149**(4–6), 341–347 (1998).
20. G. Huyet, J. K. White, A. J. Kent, S. P. Hegarty, J. V. Moloney, and J. G. McInerney, "Dynamics of a semiconductor laser with optical feedback," Phys. Rev. A **60**(2), 1534–1537 (1999).
21. F. Rogister, P. Mégret, O. Deparis, and M. Blondel, "Coexistence of in-phase and out-of-phase dynamics in a multimode external-cavity laser diode operating in the low-frequency fluctuations regime," Phys. Rev. A **62**(6), 061803 (2000).
22. I. V. Koryukin and P. Mandel, "Dynamics of semiconductor lasers with optical feedback: Comparison of multimode models in the low-frequency fluctuation regime," Phys. Rev. A **70**(5), 053819 (2004).
23. K. Yamasaki, K. Kanno, A. Matsumoto, K. Akahane, N. Yamamoto, M. Naruse, and A. Uchida, "Fast dynamics of low-frequency fluctuations in a quantum-dot laser with optical feedback," Opt. Express **29**(12), 17962–17975 (2021).
24. A. Uchida, Y. Liu, I. Fischer, P. Davis, and T. Aida, "Chaotic antiphase dynamics and synchronization in multimode semiconductor lasers," Phys. Rev. A **64**(2), 023801 (2001).
25. Y. Liu and P. Davis, "Adaptive mode selection based on chaotic search in a Fabry-Perot laser diode," Int. J. Bifurcation Chaos **8**(8), 1685–1691 (1998).
26. R. M. Nguimdo, G. Verschaffelt, J. Danckaert, and G. Van der Sande, "Reducing the phase sensitivity of laser-based optical reservoir computing systems," Opt. Express **24**(2), 1238–1252 (2016).
27. A. Bogris, C. Mesaritakis, S. Deligiannidis, and P. Li, "Fabry-Perot lasers as enablers for parallel reservoir computing," IEEE J. Sel. Top. Quantum Electron. **27**, 7500307 (2021).
28. K. Kanno, A. Uchida, and M. Bunsen, "Complexity and bandwidth enhancement in unidirectionally coupled semiconductor lasers with time-delayed optical feedback," Phys. Rev. E **93**(3), 032206 (2016).
29. R. Iwami, K. Kanno, and A. Uchida, "Chaotic mode-competition dynamics in a multimode semiconductor laser with optical feedback and injection," Opt. Express **31**(7), 11274–11291 (2023).
30. S. K. Hwang and J. M. Liu, "Dynamical characteristics of an optically injected semiconductor laser," Opt. Commun. **183**(1–4), 195–205 (2000).
31. S. Wieczorek, B. Krauskopf, and D. Lenstra, "Mechanisms for multistability in a semiconductor laser with optical injection," Opt. Commun. **183**(1–4), 215–226 (2000).
32. A. Uchida, Optical Communication with Chaotic Lasers: Applications of Nonlinear Dynamics and Synchronization (Wiley-VCH, 2012).
33. J. Ohtsubo, Semiconductor Lasers: Stability, Instability and Chaos, ed. 4 (Springer, 2017).
34. S.-J. Kim, M. Aono, and M. Hara, "Tug-of-war model for the two-bandit problem: Nonlocally-correlated parallel exploration via resource conservation," Biosystems **101**(1), 29–36 (2010).
35. S.-J. Kim and M. Aono, "Amoeba-inspired algorithm for cognitive medium access," NOLTA **5**(2), 198–209 (2014).